# Aerodynamic Loads Alteration by Gurney Flap on Supercritical Airfoils at Transonic Speeds

Amir Saman Rezaei.*

*Mechanical and Aerospace Engineering, University of California Irvine, Irvine, CA, 92617, United states*

Seyed Mohammad Jafar Sobhani, † and Amir Nejat‡

*School of Mechanical Engineering, College of Engineering, University of Tehran, Tehran, Tehran, 11155-4563, Iran*

**Effects of a gurney flap were numerically investigated on the supercritical NASA airfoil by solving the two-dimensional Reynolds-averaged Navier–Stokes equations for a range of transonic Mach numbers and angles of attack, using turbulence compressible *K - ω* SST model. The height of the gurney flap was selected to be 1.65 % chord length. A high-resolution mesh was applied to accurately predict the flow field specifically in the vicinity of the airfoil. Below the drag divergence Mach number, the gurney flap has a remarkable influence on the aerodynamic coefficients especially at -1 and 0 degrees angle of attack resulting in 50 % increase in *L/D* ratio. At high Mach numbers and angles of attack, Gurney flap loses its effects and the clean airfoil has better aerodynamic performance since it significantly boosts both the pressure and shear drag. It was observed that the gurney flap mitigates the transonic lambda shock on both surfaces of the airfoil. Moreover, it alters the Kutta condition by changing the separation point location at the trailing edge which provides the airfoil more bound circulation and lift force.**

## Nomenclature

*AOA*   =   angle of attack

*C*   =   chord length

$C_D$   =   drag coefficient

$C_L$   =   lift coefficient

---

* Ph.D. Student, UC Irvine, AIAA Student Member (723778).
† Ph.D. Candidate, School of Mechanical Engineering.
‡ Associate Professor, School of Mechanical Engineering.

| Symbol | | Description |
|---|---|---|
| $C_M$ | = | quarter-chord pitching moment coefficient |
| $C_p$ | = | pressure coefficient |
| $D_\omega$ | = | cross-diffusion term |
| $E$ | = | energy |
| $\hat{G}_K$ | = | generation of turbulence kinetic energy due to mean velocity gradients |
| $G_\omega$ | = | generation of specific dissipation rate |
| $h$ | = | enthalpy |
| $K$ | = | kinetic energy |
| $k$ | = | molecular thermal conductivity |
| $k_{eff}$ | = | effective thermal conductivity |
| $k_t$ | = | turbulent thermal conductivity |
| $L/D$ | = | lift-to-drag coefficient |
| $M$ | = | Mach number |
| $p$ | = | static pressure |
| $Re$ | = | Reynolds number |
| $u$ | = | velocity component |
| $V$ | = | overall velocity vector |
| $x, y$ | = | streamwise and normal directions |
| $Y_K$ | = | dissipation of kinetic energy due to turbulence |
| $Y_\omega$ | = | dissipation of specific dissipation rate due to turbulence |
| $\Delta C_L$ | = | increment of lift coefficient |
| $\delta_{i,j}$ | = | unit tensor |
| $\Gamma_K$ | = | effective diffusivity of kinetic energy |
| $\Gamma_\omega$ | = | effective diffusivity of specific dissipation rate |
| $\mu$ | = | molecular viscosity |
| $\omega$ | = | specific dissipation rate |
| $\rho$ | = | density |
| $\tau_{eff}$ | = | effective stress tensor |

# I. Introduction

UTILIZING high-lift devices to enhance the aerodynamic performance of the airfoils has been an active research area in the field of applied aerodynamics. Flaps, type of high-lift devices, are generally used to enhance the aerodynamic performance of the wing sections during flight. Among flap types, gurney flap is especially studied by researchers [1-5]. Gurney flap usage has opened its way to the transonic application, and transonic researchers have become interested in the influences of it on the aerodynamic characteristics of the foil which typically is the lift force augmentation. Advantages of the increased lift can lead to greater payload capacity and shorter takeoff distance. Besides, rising the lift-to-drag ratio has extra important benefits which are longer range and fuel consumption reduction.

Dan Gurney, the race car driver, was the first one who used a simple flap fixed to the top of the spoiler of his car to increase the adhesion of the tires during acceleration and cornering. Liebeck studied the effect of gurney flap on lift enhancement of Newman airfoil in subsonic region experimentally. He detected that for a particular angle of attack, the lift is increased, and the drag is decreased and concluded that for obtaining the best aerodynamic characteristics, the height of the GF must be optimized that generally lies between 1 % and 2 % of the chord length [6].

At transonic speed, one way to enhance the airfoil performance is to labor the optimization techniques [7] which have reached their point of maturity for geometry optimization. However, it is possible to improve the transonic airfoil aerodynamic performance via simple trailing edge modification, i.e. diverting the trailing edge or utilizing flaps. Computational simulations were performed by B.E. Thompson and R.D. Lotz on a supercritical airfoil [8]. The results showed when the trailing edge is divergent, the upper surface shock shifted to the downstream, increasing the *L/D*. However, gurney flap has more influence on the performance of the airfoil compared to divergent trailing edge [9], and it is a simple device that its application does not require any change in the basic shape of the airfoil. Y.C, Li conducted an experimental investigation of the effect of divergent trailing edge and gurney flap [10]. He found that in comparison with divergent trailing edge, gurney flap had a significant effect on lift and *L/D* enhancement. T. Yu et al. performed a numerical investigation on the RAE2822 airfoil and studied the Gurney flap effect at subsonic and transonic speeds by changing the height of GF and the angle of attack at a particular Mach number [11]. They concluded that the greater height of the GF generally leads to greater lift increase. However, the larger maximum lift-to-drag ratio was obtained when the height of GF is 0.25 % chord length.

It has been addressed that gurney flap affects the aerodynamics loads by changing the Kutta condition. K. Richter and H. Rosemann perused the trailing-edge devices at a transonic speed at $M = 0.775$ experimentally. The results have shown that the trailing-edge devices increase the circulation of the airfoil leading to lift enhancement and more negative pitching moment as well as an increase in minimum drag compared to the baseline configuration [12]. Manish K. Singh et al. Furthermore, Navier-Stokes analysis [13] of NACA4412 and NACA 0011 with GF indicated the similar results as [12]. Fully turbulent CFD simulation with $K - \omega$ model by Neung-SooYoo [14] showed that the Gurney flap serves to increase the effective camber of the airfoil.

One of the main applications of the Gurney flap at subsonic regions is for recovering the "lost-lift" (i.e. stall phenomena) due to variable droop leading edge [VDLE] in helicopters. Chandrasekhara [15] optimized the GF height and concluded that a 1 %-chord height flap was the most satisfactory. Maughmer [16] indicated that the drag penalty of a Gurney flap largely depends on the size of the flap. It means that increasing its size has a progressively higher drag rise. Various studies have been performed to optimize the size of Gurney flap, showing the effective height of the GF is between 1 % and 2 % chord length. Giguere et al. [17] found the best $L/D$ occurs where the height of GF was equal to the boundary-layer thickness.

Experimental investigation by Yachen li et al. [18] on a NACA 0012 airfoil at the Reynolds number of 2 million has elucidated the influences of the mounting angle and location on the effects of GF. It was concluded that in all mounting angles, there was a lift enhancement and a drag penalty. Considering those effects, the best performance was obtained in 45 degrees, but the maximum lift happens in 90 degrees. Like previous papers, the best $L/D$ ratio occurred when the GF installed at the end of the trailing edge. Perforated Gurney-Type flaps were experimented by T. Lee in a low-speed wind tunnel [19]. It was concluded that in comparison with the solid gurney, perforated gurney flaps noticeably reduce the wake size and unsteadiness results in decreasing the maximum lift and a less negative pitching moment. Also, Xuan Zhang et al. [20] demonstrated that the slit gurney flap helps in reducing the pressure drag.

It is worth mentioning that the notion of the micro-air-vehicles in the recent decades, motivated the researchers to explore the gurney flap effects on the flapping wing. Since the suction side of the foil alters during the flapping, adaptive gurney flap shows a superior performance compared to the fixed gurney flap to one side in this case [21] which makes the problem appealing for the low Reynolds number flapping researches.

All those papers conducted their simulations or experiments on specific Mach numbers. However, a transonic aircraft generally flies in a range of Mach numbers from 0.72 to 0.84. Since the performance of a supercritical airfoil varies dramatically with increasing Mach number, especially close to drag divergent Mach number, it is necessary to investigate the Mach number effects on the Gurney flap performance. In the present article, by selecting a specific height for the Gurney flap (1.65 % of the chord length) the influence of the angle of attack and the Mach number on aerodynamic coefficients are investigated. Moreover, since no informative explanation of the flow events near the trailing edge of an airfoil with gurney flap by using CFD results has been made in the previous numerical studies, the reason that how the gurney flap increases the lift by altering the Kutta condition has been addressed.

## II. Methodology

The governing equations for computations of a two-dimensional steady turbulent compressible flow, in absence of gravitational body force, external body force and any volumetric heat sources, are:

- Continuity equation

$$\frac{\partial}{\partial x_i}(\rho u_i) = 0 \qquad (1)$$

- Momentum equation

$$\frac{\partial}{\partial x_j}(\rho u_i) + \frac{\partial}{\partial x_j}(\rho u_i u_j) = -\frac{\partial p}{\partial x_i} + \frac{\partial}{\partial x_j}\left[\mu\left(\frac{\partial u_i}{\partial x_j} + \frac{\partial u_j}{\partial x_i} - \frac{2}{3}\delta_{i,j}\frac{\partial u_l}{\partial x_l}\right)\right] + \frac{\partial}{\partial x_j}\left(-\rho \overline{u_i' u_j'}\right) = 0 \qquad (2)$$

- Energy equation

$$\vec{\nabla}.[V(\rho E + p)] = \vec{\nabla}.\left[k_{eff}\vec{\nabla}T + (\tau_{eff}.\vec{V})\right] \qquad (3)$$

where $k_{eff}$ is the effective conductivity ($k + k_t$) and $\tau_{eff}.\vec{V}$ is viscous dissipation. In equation (3),

$$E = h - \frac{p}{\rho} + \frac{|\vec{V}|^2}{2} \qquad (4)$$

- Turbulence $K$ - $\omega$ SST model:

$$\frac{\partial}{\partial x_i}(\rho K u_i) = \frac{\partial}{\partial x_j}\left(\Gamma_K \frac{\partial K}{\partial x_j}\right) + \hat{G}_K - Y_K \qquad (5)$$

$$\frac{\partial}{\partial x_j}(\rho \omega u_j) = \frac{\partial}{\partial x_j}\left(\Gamma_\omega \frac{\partial \omega}{\partial x_j}\right) + G_\omega - Y_\omega + D_\omega \qquad (6)$$

The $K$ - $\omega$ SST model has been used as it has shown promising results for steady and unsteady aerodynamic simulation [11, 20, 24]. The simulation was carried out using the implicit method, and all the discretization was set to be second-order upwind. The convergence criteria were to require a normalized residual less than $10^{-6}$ for all equations.

Fluent version 6.3 was used as the CFD package solver in the present study. Settings in the CFD software were:

1) Selecting the ideal gas model for air density change.

2) Choosing turbulent $K$ - $\omega$ SST model as the viscous model and retaining the default values parameters.

3) Enabling energy equation.

4) Set Mach number and angle of attack for pressure far-field boundary condition.

In all the plots and tables, "OR" stands for the original airfoil and "GF" stands for the airfoil with gurney flap attached to it.

## A. Geometry Modeling and Grid Generation

Supercritical airfoil SC(2)-0412 [25] with the chord length of 1 m is surrounded by a freestream which its Mach number differs from 0.66 to 0.84. C-type grid (Fig. 1a) was used for meshing the domain. The radius of the C part of the domain is 20 times larger than the chord length and far field length in downstream is the same. In order to generate the mesh, 250 nodes are specified at each side of the airfoil that are denser in the vicinity of the leading and trailing edges. The distance of the first grid from the wall surface in the boundary layer is set to $5\times10^{-5}$ m with the growth ratio of 1.18. The total number of grids is almost 200,000. With this mesh topology, not only more reliable results are expected for aerodynamic coefficients especially for $C_D$, but also vortices can be illustrated more properly. In this study, the thickness of the trailing edge is 0.55 % of chord length and since in almost all the papers demonstrated that the optimum length of gurney flap should be between 1 % and 2 % chord length, therefore height of gurney flap for the present airfoil is selected 3 times larger than length of trailing edge or in the other words it is equal to 1.65 % chord length (Fig. 1b).

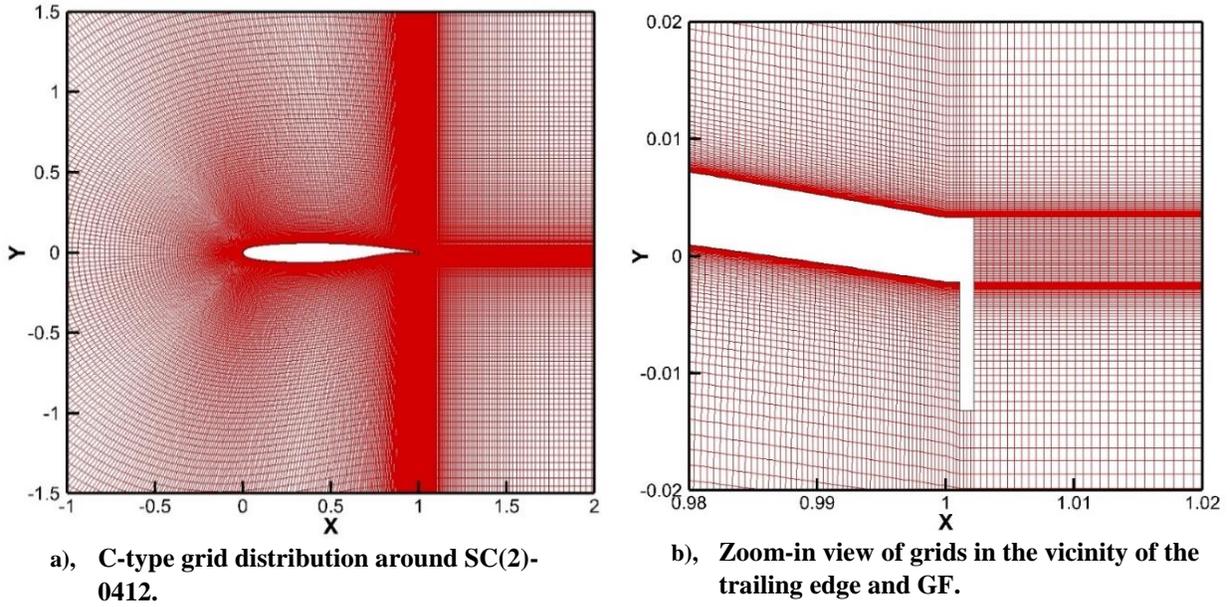

| a), C-type grid distribution around SC(2)-0412. | b), Zoom-in view of grids in the vicinity of the trailing edge and GF. |

**Fig. 1 Grid System.**

## III. Validation Study

To verify both the simulation methodology and the grid resolution used in the present study, the supercritical RAE 2822 airfoil, case 9 of AGARD report [22], simulated with the same grid resolution and the solution method in transonic flow as described in the previous section. The flow conditions of the test case are: $M = 0.729$, $AOA = 3.19º$ and $Re = 6.5 \times 10^6$. In order to compare the results with the wind tunnel experimental results, the condition of the flow in the numerical simulation should be corrected to compensate for the wall effects of the wind tunnel. Using the correction method of [23], Mach number and $AOA$ is set to be 0.729 and 2.79º respectively, and Re = 6.5×106. Table 1 shows the aerodynamic coefficients of the present simulation by $K$ - $\omega$ SST model in comparison with the AGARD report. Fig. 2 shows that the obtained pressure distribution from the numerical solution agrees well with the experimental results.

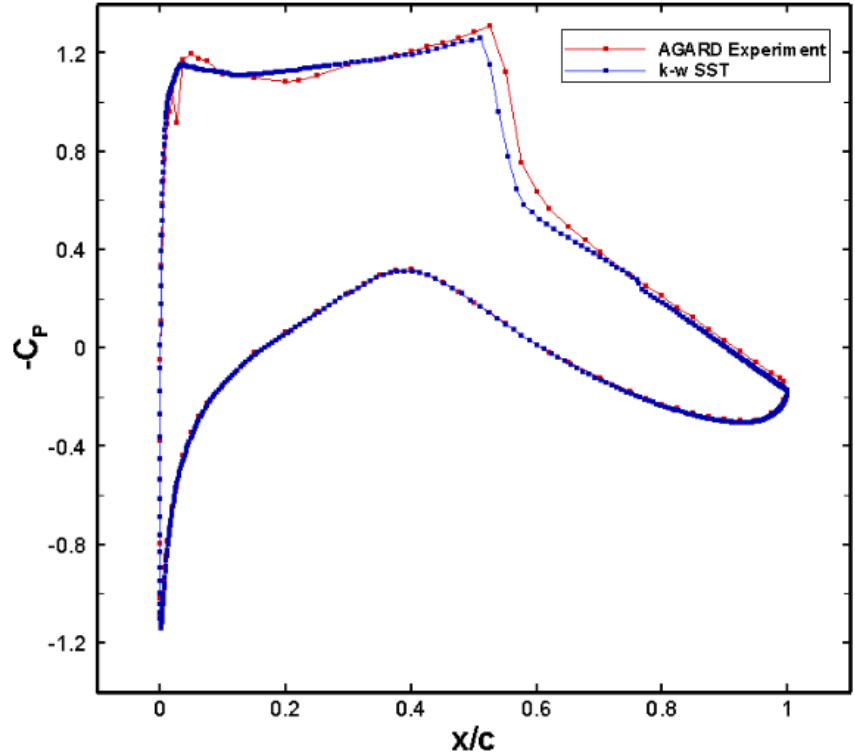

**Fig. 2  Comparison of pressure distribution between the AGARD experiment and $K$ - $\omega$ SST simulation.**

**Table 1 Comparison of AGARD report and CFD simulation**

| Report | $C_L$ | $\Delta C_L$ | $C_D$ | $\Delta C_D$ | L/D | $\Delta(L/D)$ | $x_{shock}/C$ |
|---|---|---|---|---|---|---|---|
| AGARD | 0.803 | - | 0.0168 | - | 47.79 | - | 0.525 |
| CFD | 0.781 | 2.7 % | 0.0170 | 1.2 % | 45.94 | 3.8 % | 0.52 |

## IV. Results and Discussion

Here we investigate the effect of Mach number (at transonic speeds) on the performance of the gurney flap in detail. The current issue has not been addressed in the literature appropriately based on the authors' knowledge. The simulations were carried out for the far filed Mach numbers of 0.66 up to 0.84 with 0.02 incremental Mach number steps, and the angle of attack varies between -2 to 3 degrees with 1-degree increment.

### A. Effect of gurney flap on the Lambda shock at the transonic speed

To understand the effect of the gurney flap on the aerodynamic characteristics of the airfoil, a comparison between the OR and the GF airfoil was performed in terms of pressure distribution. According to the pressure coefficient diagram, it can be interpreted from the sudden changes in the pressure that in the original case at zero *AOA*, there is a shock on the upper surface from $M = 0.78$ (Fig. 3a) and on the bottom surface from $M = 0.82$ (Fig. 3b). These strong lambda shocks on the surfaces of the foil originated from the transonic inlet flow, noticeably influences the exerted forces and moments on the body.

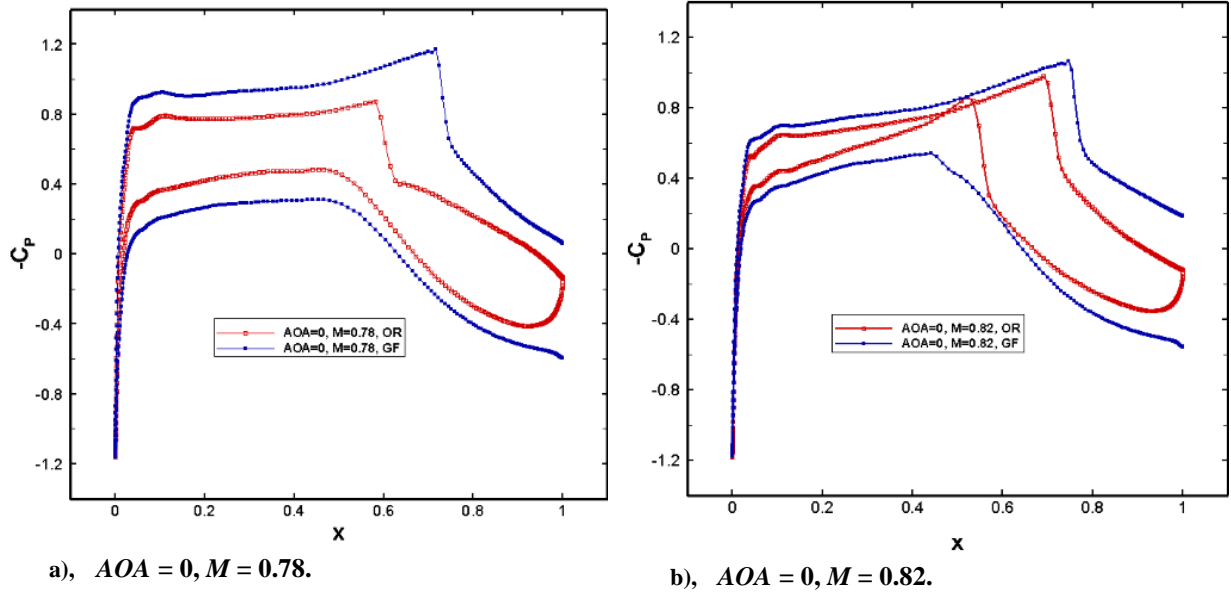

a), $AOA = 0, M = 0.78.$  b), $AOA = 0, M = 0.82.$

**Fig. 3 Pressure coefficient distribution for Original and GF airfoil.**

Utilizing Gurney flap delays the transonic shock formation on the upper surface of the airfoil. The difference between pressure coefficient distribution for clean airfoil and airfoil with GF is shown in [Fig. 3a] for flow condition of $M = 0.78$ and $AOA = 0$. Increase in the pressure differential between lower and upper surfaces for the GF airfoil against the OR one demonstrates the lift enhancement. As it is seen, the position of the transonic shock has been pushed back from 0.58C in clean airfoil to 0.7C in GF case; it means that there is about 21 % put-off ratio in the position of the transonic shock. The same scenario happens when a lambda shock forms on the lower surface (Fig. 3b). For better illustration in this case where the Mach number at far-field is 0.82, the Mach number contours are provided in Fig. 4. The lambda shocks are pointed with black arrows. In this case, the gurney flap efficacy is even more sever on the flow physics as it has attenuated the shock on the lower surface and delayed the shock on the upper surface ensuing the pressure loss recovery compared to the OR case. Altogether, in the regimes where the shock happens on one side of the foil (typically on the suction side), gurney flap postpones the shock formation drastically and help in increasing the pressure differential with the other side. If the Mach number is high enough that a secondary shock occurs on the pressure side as well, gurney flap cope with attenuation of the shocks on both sides as if its aid is proportionally divided to each side.

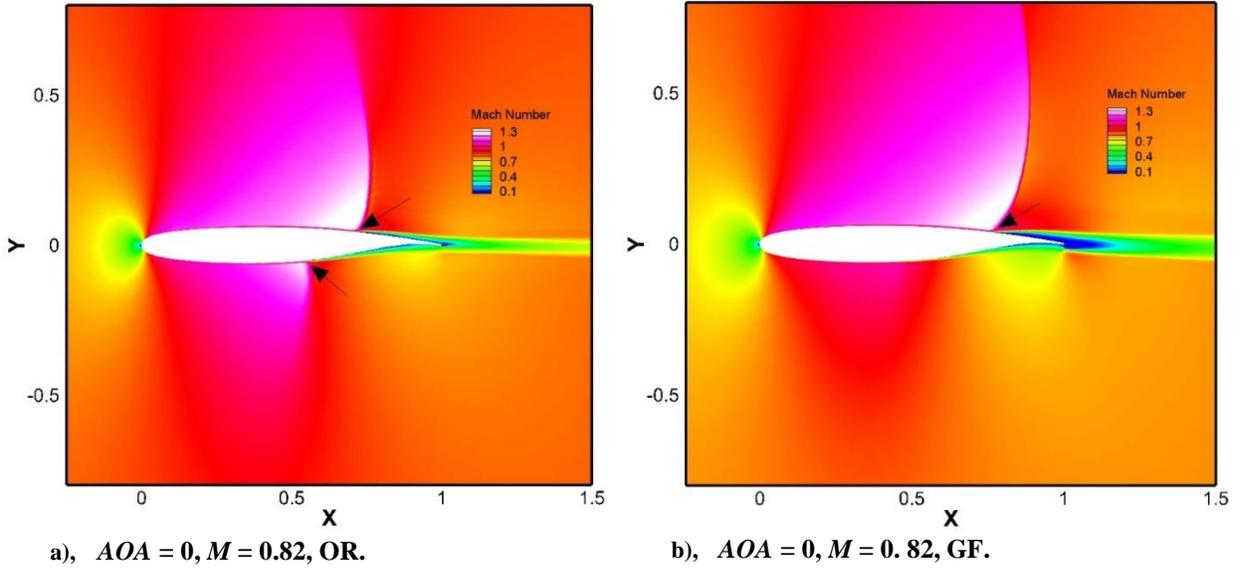

**a), $AOA = 0, M = 0.82$, OR.**  **b), $AOA = 0, M = 0.82$, GF.**

**Fig. 4 Mach number contour near the airfoil at $M = 0.82$ that shows the Lambda shock on the surface.**

Fig. 5 and Fig. 6 indicate the effect of gurney flap on the upper surface shock in a range of angle of attacks and Mach numbers. Results are also tabulated in Table 2 and Table 3. As it is shown, by increasing either *AOA* or Mach number, shock location put-off ratio decreases. In other words, when the shock becomes stronger, gurney flap influence becomes weaker. Consequently, in those conditions using gurney flap is not recommended. This subject will be discussed more in the next figures.

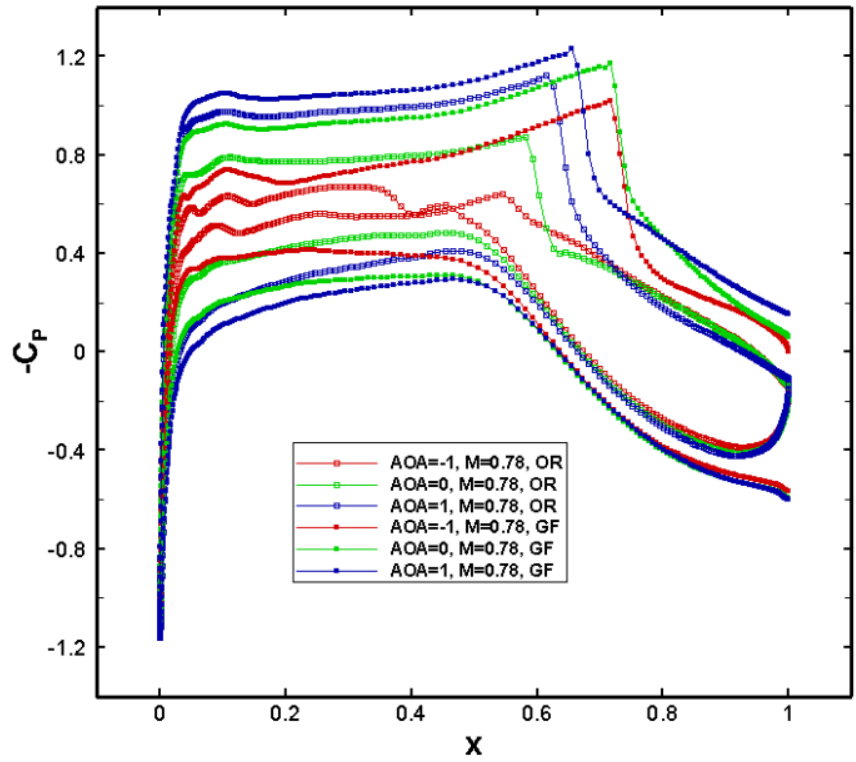

**Fig. 5 Transonic shock at $M = 0.78$.**

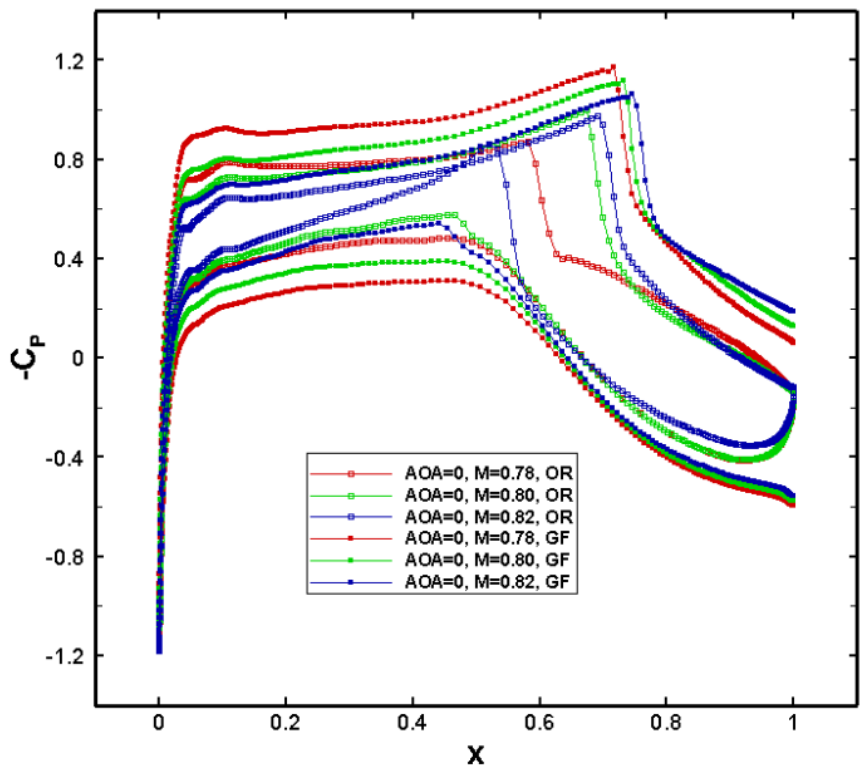

**Fig. 6 Transonic shock at $AOA = 0$.**

**Table 2 Upper surface shock place at $M = 0.78$**

| AOA | $x_{shock-OR}/C$ | $x_{shock-GF}/C$ | Put-off ratio |
|---|---|---|---|
| -1° | 0.55 | 0.72 | 31 % |
| 0 | 0.58 | 0.71 | 22 % |
| 1° | 0.63 | 0.68 | 8 % |

**Table 3 Upper surface shock place at $AOA = 0$**

| M | $x_{shock-OR}/C$ | $x_{shock-GF}/C$ | Put-off ratio |
|---|---|---|---|
| 0.78 | 0.58 | 0.71 | 22 % |
| 0.80 | 0.68 | 0.75 | 10 % |
| 0.82 | 0.7 | 0.76 | 8 % |

**B. Effect of varying Mach number and angle of attack on the performance of the OR and GF airfoils**

Fig. 7, Fig. 8, and Fig. 9 display the aerodynamic force coefficients as well as the quarter-chord pitching moment coefficient for all the cases versus Mach numbers in the various angle of attacks.

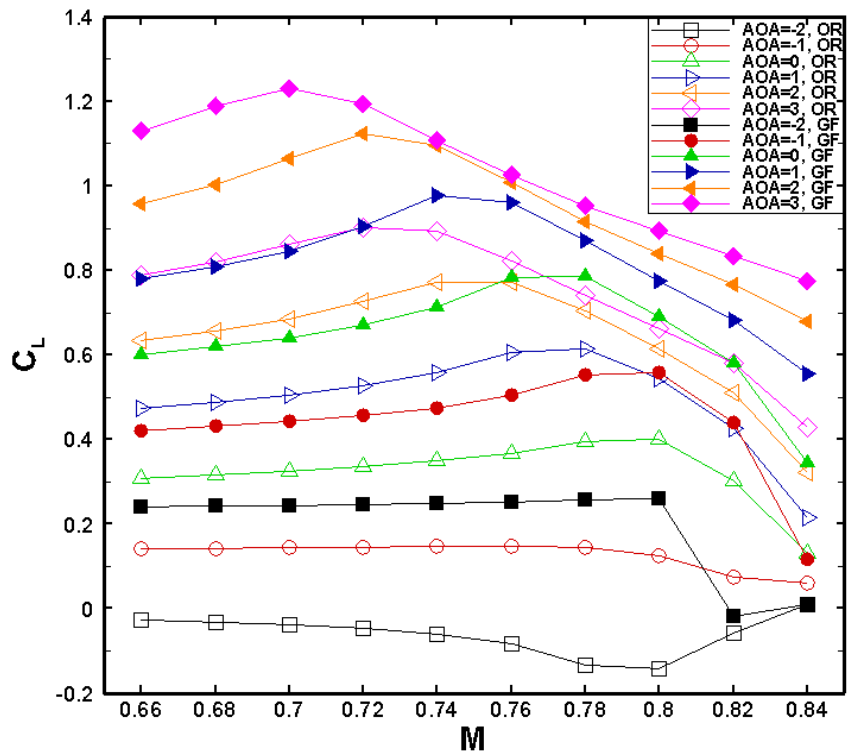

**Fig. 7 Lift coefficient vs. Mach number for the original airfoil and GF case.**

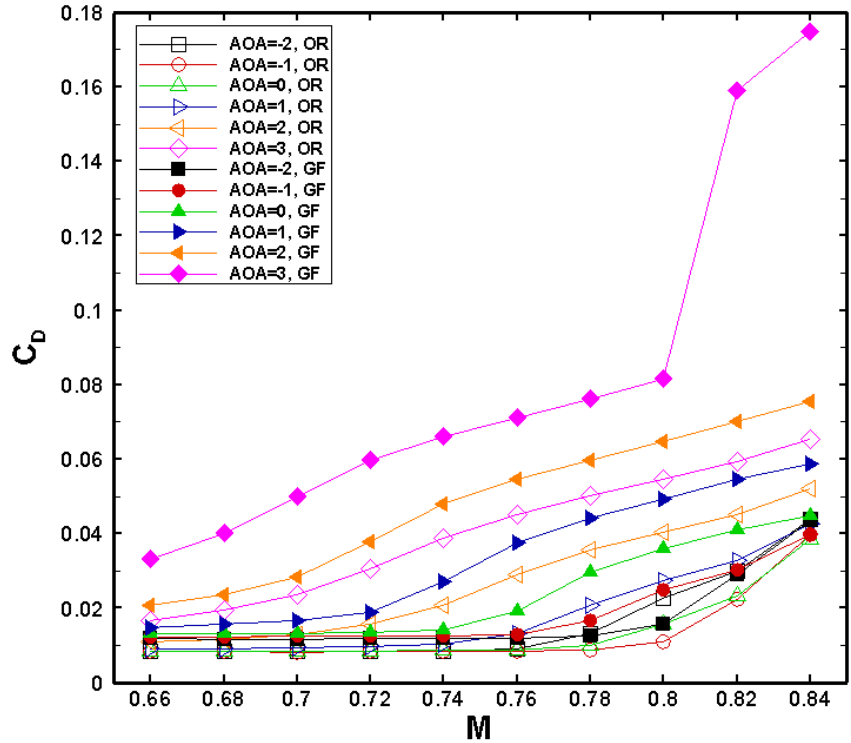

**Fig. 8  Drag coefficient vs. Mach number for the original airfoil and GF case.**

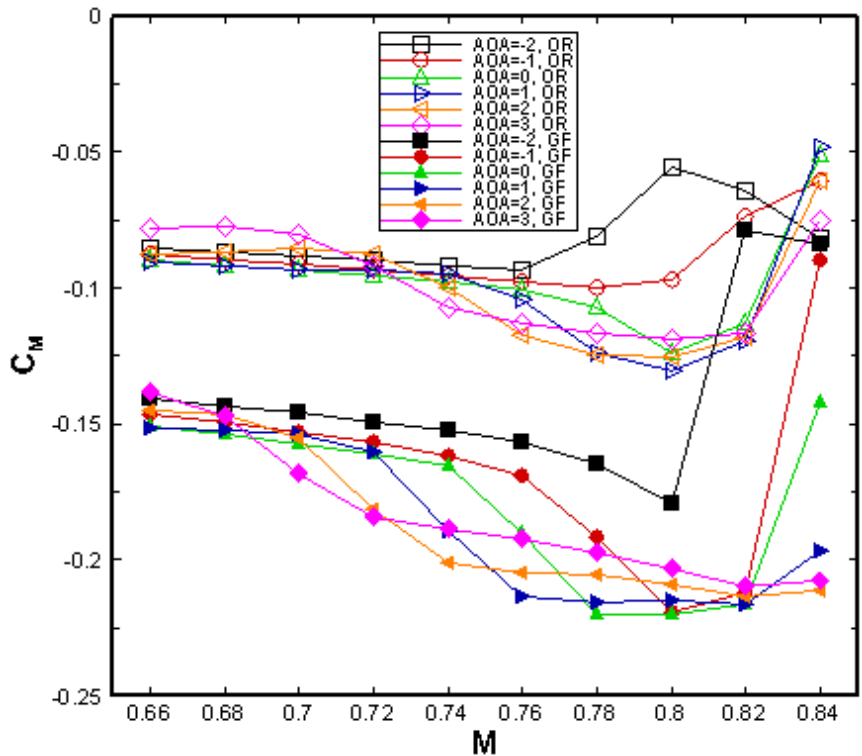

**Fig. 9  Quarter-chord pitching moment coefficient vs. Mach number for the original airfoil and GF case.**

Since there are two variables in this study, Mach number and angle of attack, results can be interpreted as 3-dimensional figures instead of 2D. Surfaces of drag, lift, lift to drag ratio and pitching moment coefficients are shown in Fig. 10. For the better schematic, $C_M$ figures were illustrated with the positive sign.

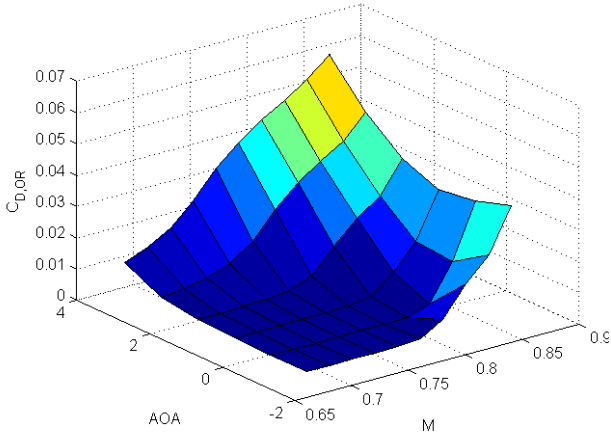

a), Drag coefficient surface for the original airfoil.

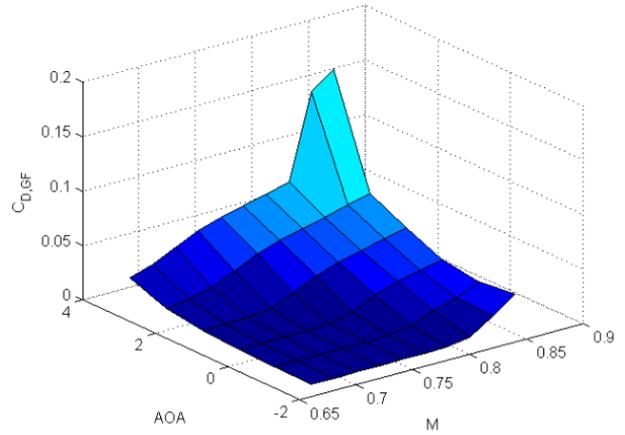

b), Drag coefficient surface for GF case.

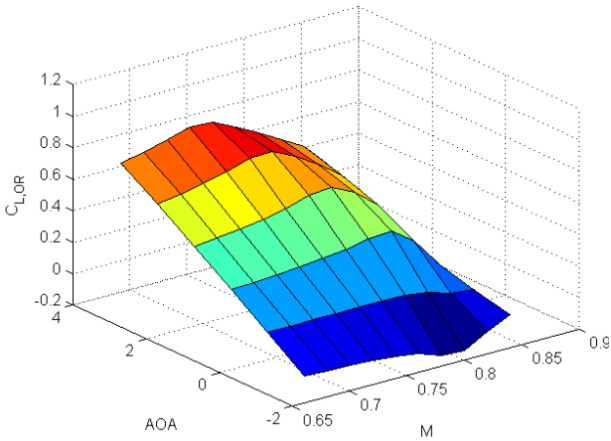

c), Lift coefficient surface for the original airfoil.

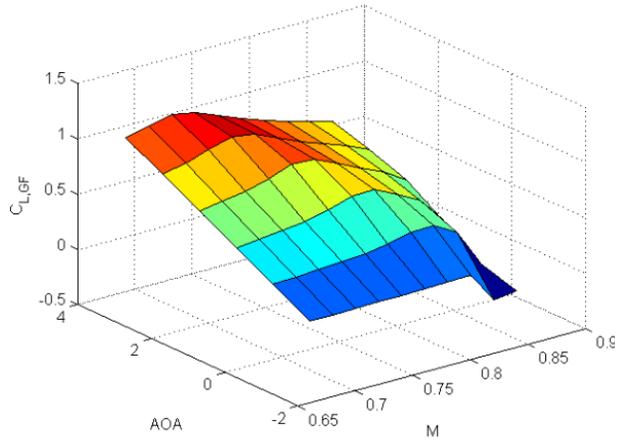

d), Lift coefficient surface for GF case.

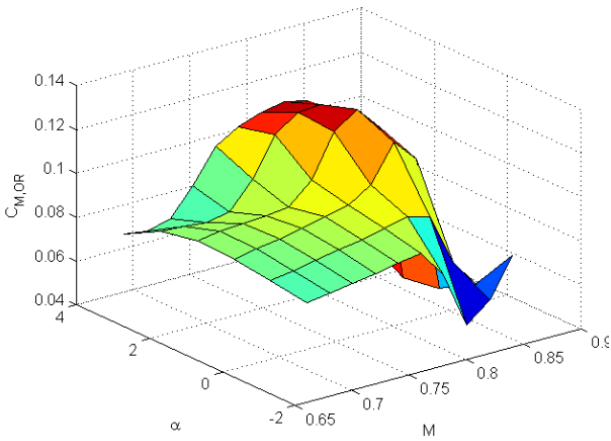 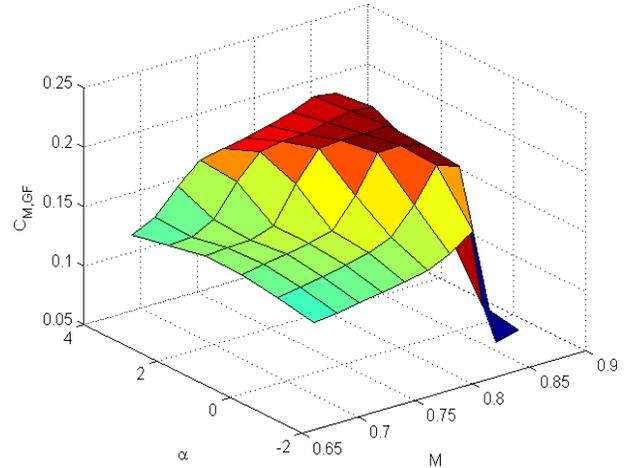

e), **Quarter-chord pitching moment coefficient surface for the original airfoil.**

f), **Quarter-chord pitching moment coefficient surface for GF case.**

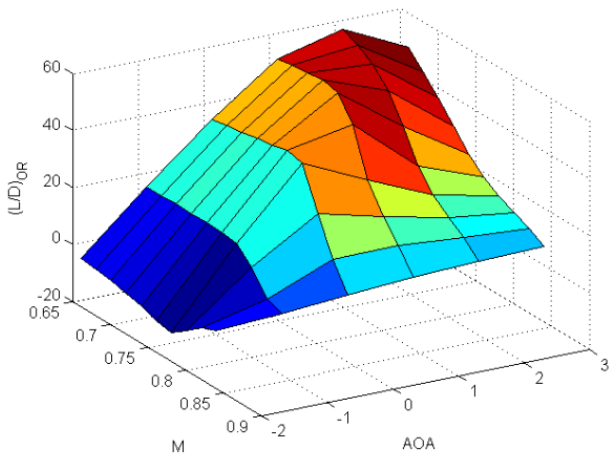 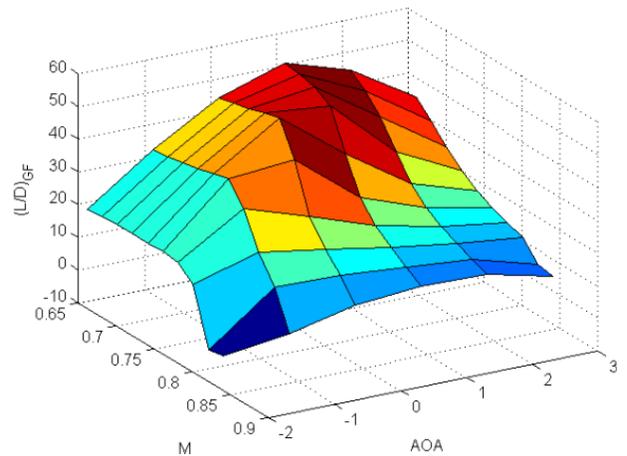

g), *L/D* **ratio surface for the original airfoil.**

h), *L/D* **ratio surface for GF case.**

**Fig. 10    3D surfaces of aerodynamic coefficients.**

One important impact of Gurney flap is identified in 3D surfaces of aerodynamics coefficients (Fig. 10). Regardless of fall or rise in the amount of those coefficients, GF airfoil surfaces are smoother and have less sharp variation in comparison with the original airfoil, making the airfoil more reliable and stable as the flight conditions change.

In general, increasing in lift coefficient is obtained in all the studied Mach numbers and *AOA*s compared to the original. This impact decreases gradually by increasing Mach number; for example, at *AOA* = -1º, 0, 1º, $C_L$ is increased by at least 1.5 times. The zero-lift angle of attack is smaller for GF type airfoil. Therefore, the similar lift in GF type airfoil is attainable in a lower angle of attack compared to the original airfoil. This can lead to lower flight angle of attack for the complete aircraft reducing the total body drag of the aircraft. For instance, in GF case at *AOA* = -1º the lift coefficient is still more than $C_L$ at $AOA = 0$ in the original case about 1.3 times as an average for a range of studied Mach numbers.

Decreasing in quarter-chord pitching moment coefficient is due to circulation enhancement and is reported in Fig. 9. This phenomenon is the result of the increment of the near body forces mostly at the aft side of the airfoil (closer to TE) which can be addressed in all the pressure coefficient plots.

At $AOA = 0$, before drag divergence Mach number, the effect of GF on lift coefficient is more remarkable than on drag coefficient resulting in *L/D* ratio rise (Fig. 11). It can be seen better performance for GF airfoil in all Mach numbers at $AOA = -1°$. At high angles of attack, the original airfoil has better *L/D* before $M = 0.8$ but at higher Mach numbers original and GF airfoils have almost the same *L/D*. Generally speaking, it cannot be concluded that using the gurney flap is beneficial for all Mach numbers and angle of attacks, but it improves the airfoil performance for several flight conditions and may be considered in designing the transonic airfoils.

Lift-to-drag ratio vs. lift coefficient is shown in Fig. 12. The maximum *L/D* ratio will not be affected by Gurney flap significantly but increase in the lift coefficient for a given lift-to-drag ratio is remarkable. The GF type airfoil has a higher lift coefficient at the same *L/D* and Mach numbers in lower angles of attack. It can be seen that in lower angles of attack in a certain *L/D*, very higher $C_L$ can be acquired by using the gurney flap. Since the designed cruise $C_L$ for transonic jetliners is between 0.45 and 0.65, by looking at *L/D* diagram versus $C_L$ (Fig. 12), it can be concluded that the best performance, as it was mentioned, differs from case to case.

The drag polars (Fig. 13) indicates that for the cruise lift coefficients, there is an increase in drag coefficient, which will be amplified by increasing the *AOA*. Furthermore, it can be seen that when the lift coefficient begins to decrease, the negative slope for decreasing lift coefficient versus drag coefficient is the same for the original airfoil and GF type airfoil. The most increment in $C_L$ exists in $AOA = -1°$ that is obvious from Fig. 14. By increasing the angle of attack, the increment in $C_L$ has a descending trend.

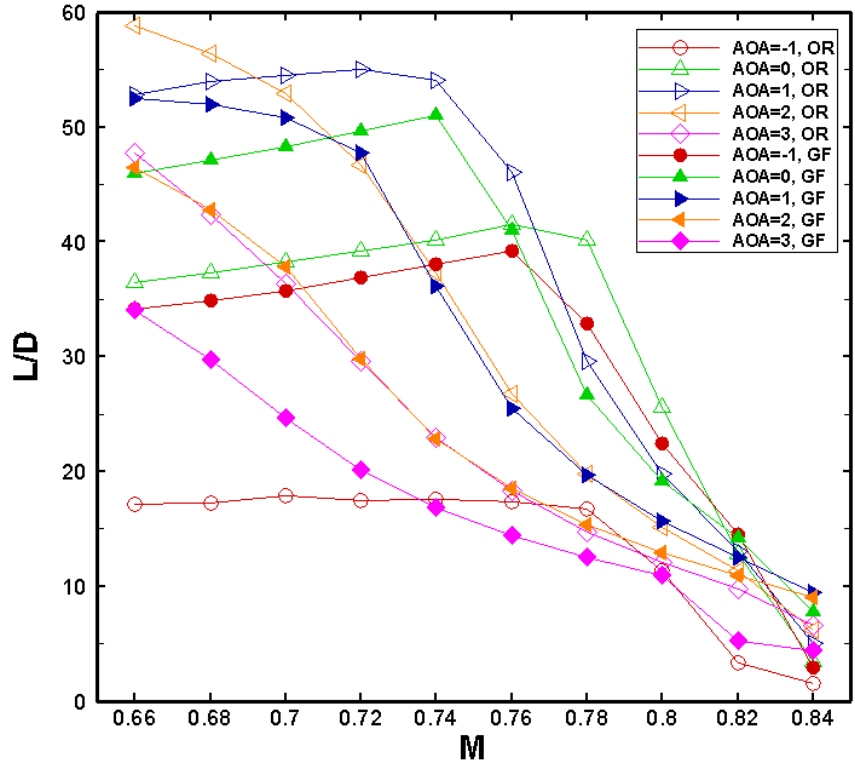

**Fig. 11** *L/D* ratio vs. Mach number for the original airfoil and GF case.

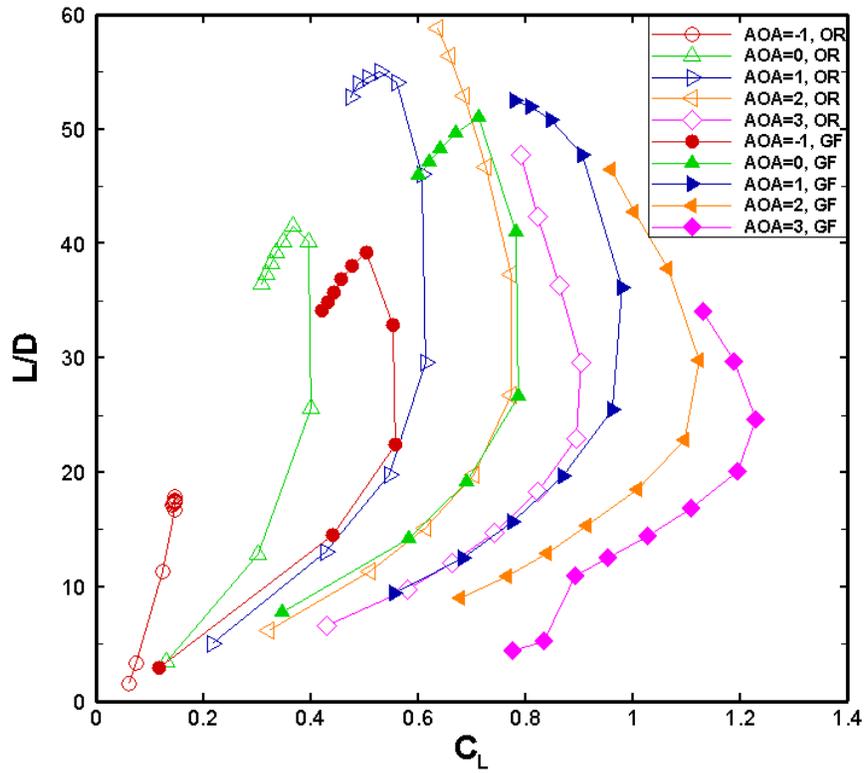

Fig. 12    *L/D* ratio vs. lift coefficient for the original airfoil and GF case.

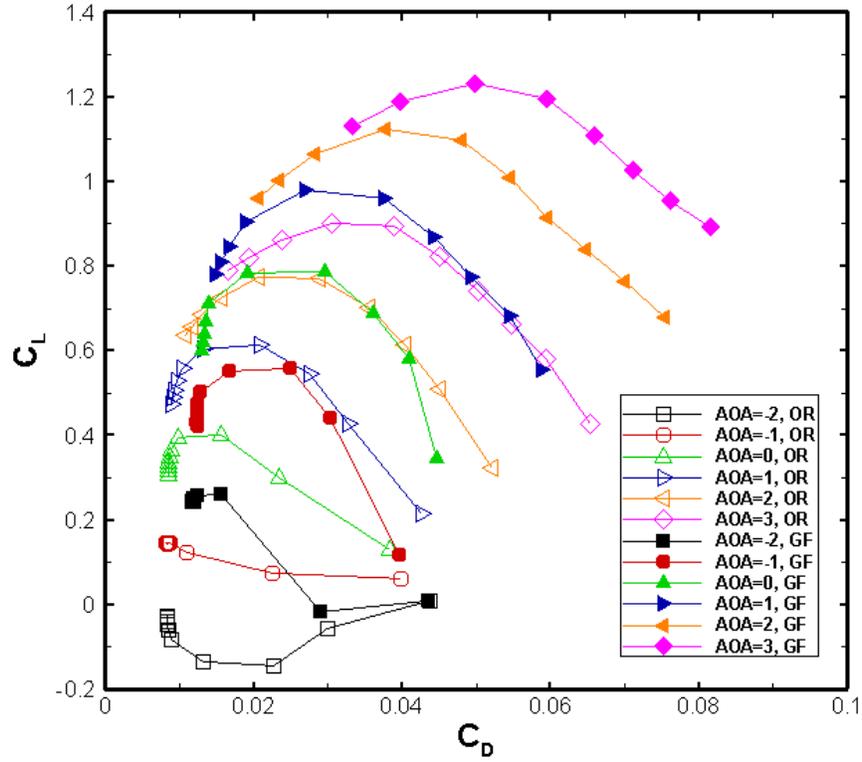

Fig. 13    Drag polar at constant *AOA* for the original airfoil and GF case.

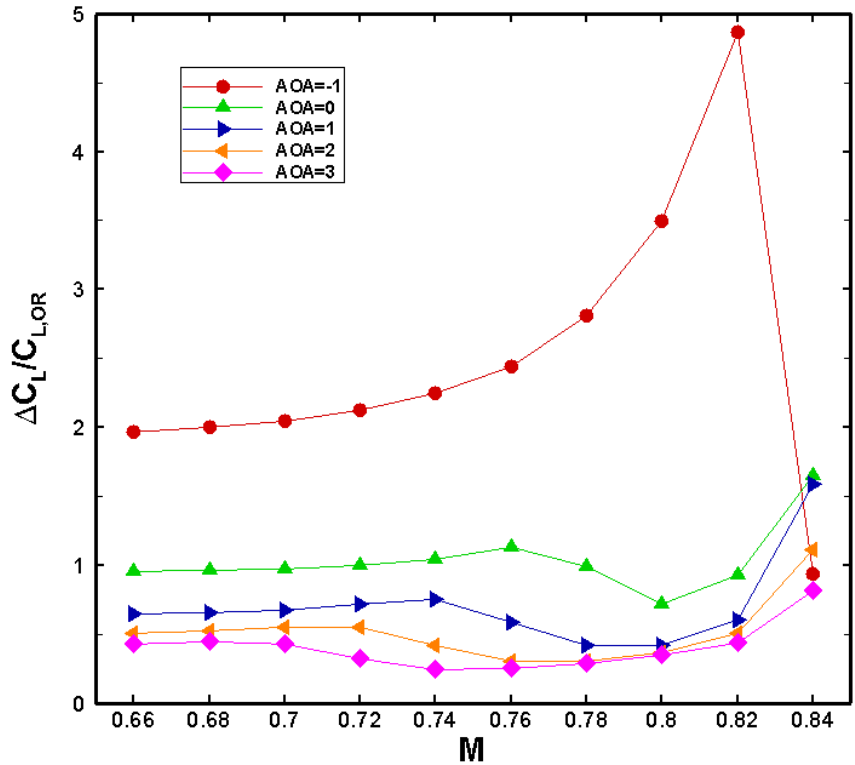

**Fig. 14** $\Delta C_L / C_{L,OR}$ **vs. Mach number.**

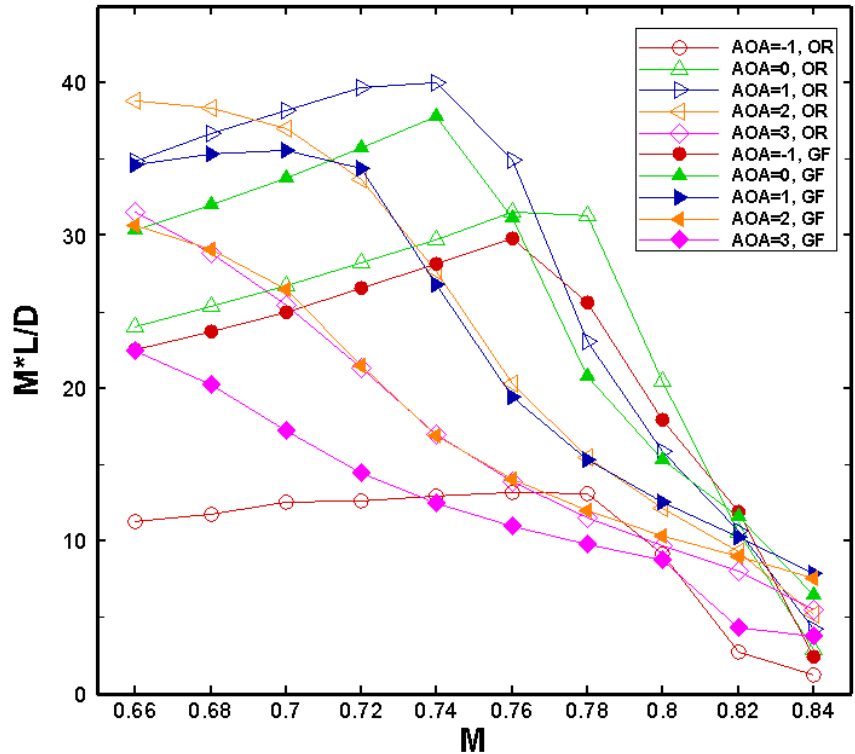

**Fig. 15** $M \times L/D$ **vs. Mach number for the original airfoil and GF case.**

Fig. 15 is another important diagram which can specify whether to use GF or not. Since $M \times L/D$ factor is the critical parameter for the flight range, the proper flight condition can be chosen for the aircraft with GF airfoil. With considering that factor, in Mach numbers less than $M = 0.8$, at $AOA = -1°$ and 0, GF airfoil has better performance, but in other cases, the original airfoil would have higher flight range. In the end, if we look at the Fig. 15 carefully, we notice that at high Mach numbers, again GF airfoil gains little better performance.

**C. Investigating the Effects of Gurney Flap on Lift and Drag Forces**

The main effect of using the gurney flap is to improve the $C_L$ that is the result of circulation enhancement by introducing two counter-rotating vortices near the trailing edge (Fig. 16a). Accordingly, the pressure is decreased on the suction side and is increased on the pressure side. To shed more light towards the function of the gurney flap, vorticity contours are presented in Fig. 16. At low angles of attack and high Reynolds number, the separation occurs at the trailing edge and the classical Kutta condition is satisfied [26]. The positive vorticity generated from the bottom surface of the airfoil and the negative one from the upper surface are shed into the wake, and the conservation of circulation law will lead to the circulation build up on the airfoil which ultimately is translated to the lift force through Kutta-Joukowski theorem. For the airfoils that possess sharp trailing edge, i.e. RAE 2822 (Fig. 16b), the coming vorticities from the two sides of the airfoil, separate at the sharp edge. As it is expected, for the airfoils with a blunt trailing edge like the selected airfoil for the current study, the aforementioned scenario happens almost at the middle of the blunt edge (Fig. 16d). For both cases, the camberline passes through this point. Nevertheless, adding the gurney flap at the trailing edge will shift the separation point toward the bottom of the gurney flap (Fig. 16c). Two interpretation of this aspect is possible. First, shifting the separation point downward makes the airfoil more cambered which in turn increases the lift. Second, as it can be observed by comparing Fig. 16c and Fig. 16d, the vorticity flux to the wake is significantly increased due to the flap that augments the amount of bound circulation on the airfoil.

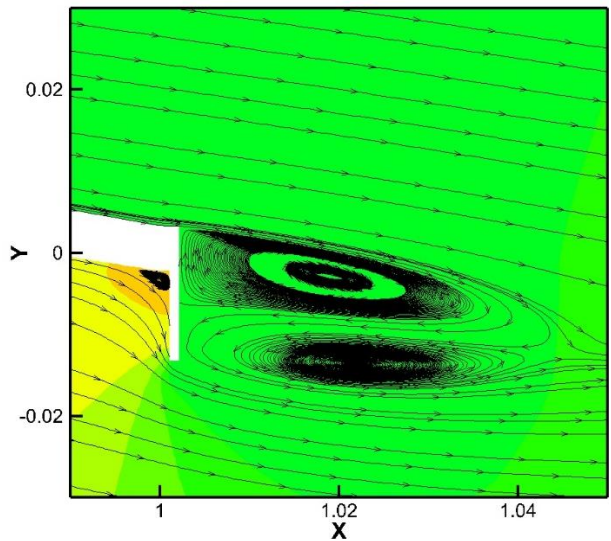
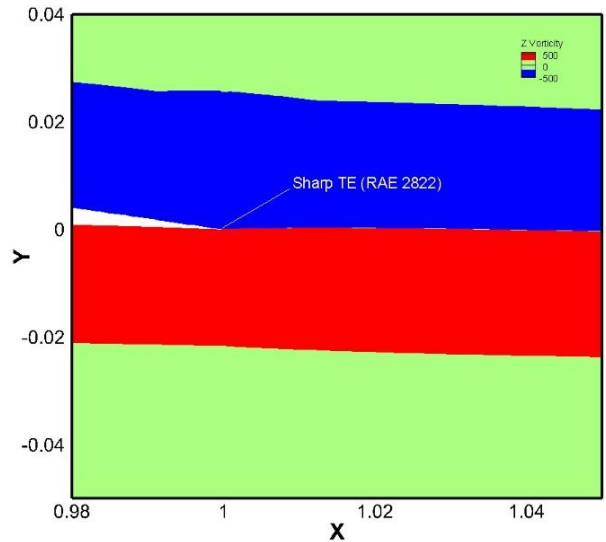

a), Counter-Rotating vortices near the TE with GF.

b), Vorticity contour for RAE 2822 at zero AOA.

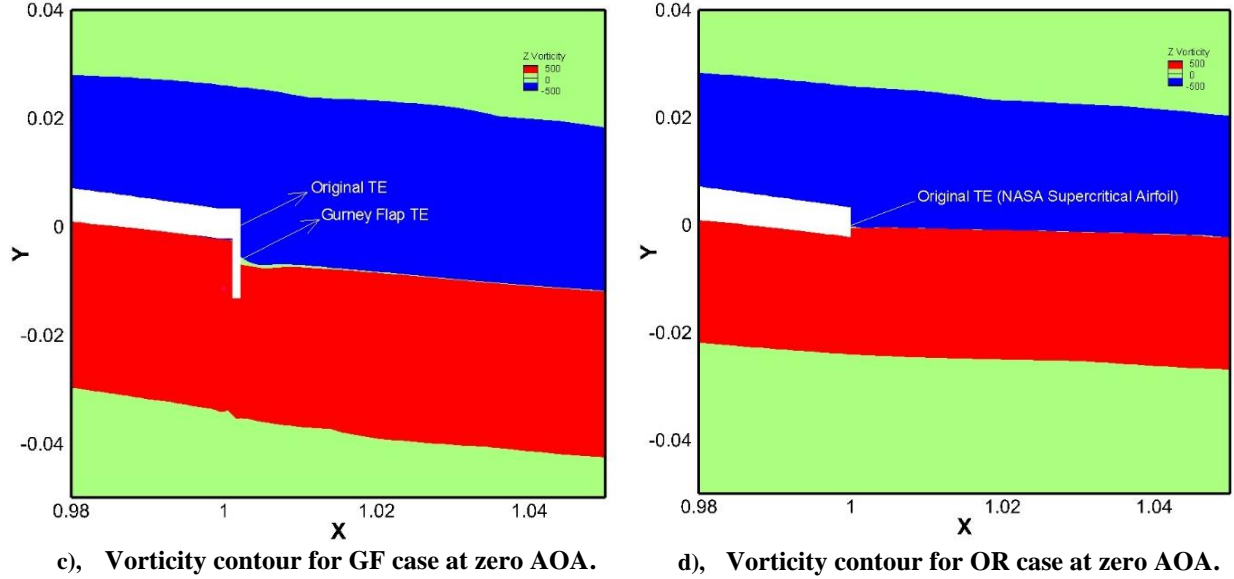

c),   **Vorticity contour for GF case at zero AOA.**     d),   **Vorticity contour for OR case at zero AOA.**

**Fig. 16    Flow physics near the trailing edge at *AOA* = 0 and *M* = 0.74.**

In order to obtain more insight into the circulation build up around the airfoil, a control volume was considered around the airfoil (Fig. 17). Its right side is aligned with the trailing edge and the dimension of the other sides was selected such that the control volume contains the boundary layer entirely and is not affected by the flow changes upstream of the foil. A considerable bigger control volume was chosen, and no difference was observed in the results. The amount of circulation was calculated by:

$$\Gamma = \oint v.dl \tag{5}$$

and is normalized by half of the product of the free stream velocity and the chord length, i.e. $\frac{1}{2}U_\infty C$. The results were calculated for four different Mach numbers and the angle of attack is zero in all the cases (Table 4). It can be seen that the gurney flap increases the circulation around the airfoil by almost 100 %. This effect is magnified at very high Mach numbers where two strong shocks sit on both surfaces of the airfoil. In this case, utilizing gurney flap can potentially increase the bound circulation by almost 160 % which is remarkable.

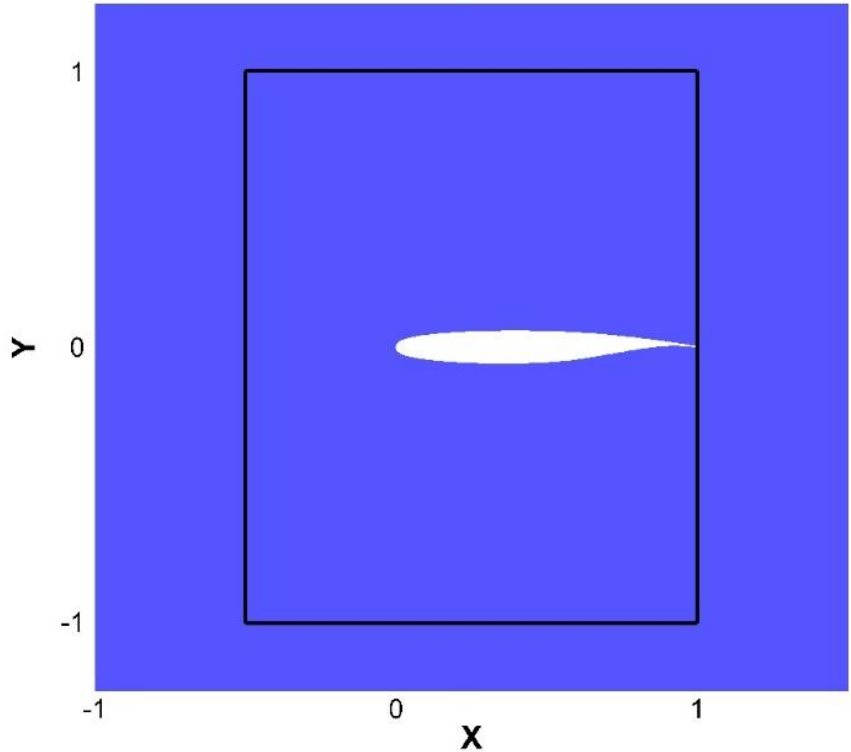

**Fig. 17 Control volume around the airfoil to calculate the bound circulation.**

**Table 4 Normalized circulation around the airfoil at *AOA* = 0**

| M | OR | GF | Increment (%) |
|---|---|---|---|
| 0.66 | 0.31 | 0.6 | 93 |
| 0.78 | 0.4 | 0.79 | 97.5 |
| 0.82 | 0.3 | 0.58 | 93.3 |
| 0.84 | 0.15 | 0.4 | 166.6 |

The higher $C_D$ that is the penalty of $C_L$ augmentation, which can be seen in Fig. 8, is the result of larger wave drag (stronger shock) and the pressure drag caused from vortices created by gurney flap in the vicinity of the trailing edge as shown in Fig. 16. In order to quantitatively analyze the effect of gurney flap on the aerodynamic coefficients, specifically drag coefficient, the two sources of applied forces on the airfoil have been integrated over the airfoil and the percentage of the contribution of each of them on the total force have been provided in Table 4. Not surprisingly, most of the lift force is produced by the pressure forces in all the cases. On the other hand, at a certain Mach number, say 0.74, the gurney flap significantly increases the contribution of the pressure force on the drag, which is the consequence of the low-pressure region due to the counter-rotating vortices near the trailing edge. This effect diminishes at higher angles of attack. Moreover, at higher Mach number where the lambda shock is vigorous, even for the original airfoil, most of the drag force comes from the pressure forces and adding the gurney flap may slightly amplify the pressure contribution. A substantial conclusion is that at lower Mach number where no sever shock is observed on the foil, the pressure drag rises remarkably by attaching the gurney flap. However, at higher Mach number where there is a huge pressure drag due to the sitting shock, the effect of the gurney flap on increasing the pressure drag contribution is negligible because of the following. At high Mach numbers, the gurney flap induces two aspects to the flow field. First is to create a suction region near the trailing edge tending to increase the pressure drag, and second is to weaken the lambda shock on the airfoil tending to decrease the pressure drag. This interaction leads to almost the same pressure drag contribution with or without the gurney flap. Note that Table 4 exhibits the percentage of the pressure and shear forces on the lift and drag for each case, not the actual value of the forces, as both the $C_D$ and $C_L$ increase by adding the gurney flap.

**Table 5 Contribution of pressure and shear forces on the aerodynamic coefficients**

| M | AOA | $C_L$ | | $C_D$ | |
|---|---|---|---|---|---|
| | | Pressure (%) | Shear (%) | Pressure (%) | Shear (%) |
| 0.74 OR | 0 | 99.99 | 0.01 | 43.5 | 56.5 |
| 0.74 GF | 0 | 99.99 | 0.01 | 63.7 | 36.3 |
| 0.74 OR | 3° | 99.99 | 0.01 | 88.7 | 11.3 |
| 0.74 GF | 3° | 99.99 | 0.01 | 94 | 6 |
| 0.84 OR | 0 | 99.99 | 0.01 | 89.6 | 10.4 |
| 0.84 GF | 0 | 99.99 | 0.01 | 90.6 | 9.4 |

## V. Conclusion

Numerical turbulence transonic simulation was performed to investigate the effect of Mach number at transonic speeds and different angles of attacks for NASA SC(2)-0412 airfoil with Gurney flap. Gurney flap increases the amount of $C_L$, $C_D$, and $C_M$ in all the studied Mach numbers and angles of attack. At $AOA = -1°$ in all Mach numbers, better aerodynamic performance is obtained by utilizing Gurney flap providing lift coefficient between 0.4 and 0.5 which is appropriate for cruise condition. A rise in the angle of attack would adversely affect the aerodynamic performance of the GF airfoil in transonic speed, especially after drag divergence Mach number. For instance, at $AOA = 0$ better $L/D$ can be achieved by installing Gurney flap before drag divergence Mach number, but after drag divergence, a rapid increase in $C_D$ is observed which decrease the $L/D$.

At transonic speeds where lambda shocks are expected on top and bottom of the foil, the gurney flap alleviates and delays the shock formation and, in some cases, eliminates them, results in a higher pressure differential between suction and pressure sides of the supercritical airfoil that dictates higher lift force. Furthermore, the gurney flap serves to increase the pressure contribution in the drag force which is more dominant at low Mach number and low angle of attack.

The lift augmentation has been attributed to the intricate flowfield near the trailing edge. Adding the gurney flap to the airfoil will bring the trailing edge separation point toward the bottom of the flap. This can be described as increasing the camber of the airfoil which results in more lift force. Also, it has been shown that the generated positive vorticity flux from the lower surface of the airfoil, increases significantly that raises the bound circulation over the airfoil.